\documentclass[letterpaper,useAMS,usenatbib]{mn2e}
\usepackage[totalwidth=500pt, totalheight=680pt]{geometry}
\usepackage{graphicx,longtable}

\title[Astrophysical counterparts of IceCube events]
{Are both BL Lacs and pulsar wind nebulae the astrophysical counterparts of IceCube neutrino
  events?}
\author[P. Padovani and E. Resconi]{P. Padovani$^{1}$ and E. Resconi$^{2}$\thanks{E-mail:
ppadovan@eso.org, elisa.resconi@tum.de}\\
$^{1}$European Southern Observatory, Karl-Schwarzschild-Str. 2, D-85748 Garching bei M\"unchen, Germany\\
$^{2}$Technische Universit\"at M\"unchen, James-Frank-Str. 1, D-85748 Garching bei M\"unchen, Germany}
\begin{document}
\newdimen\digitwidth
\setbox0=\hbox{2}
\digitwidth=\wd0
\catcode `#=\active
\def#{\kern\digitwidth}


\pagerange{\pageref{firstpage}--\pageref{lastpage}} \pubyear{2014}

\maketitle

\label{firstpage}

\begin{abstract}
IceCube has recently reported the discovery of high-energy neutrinos of
astrophysical origin, opening up the PeV (10$^{15}$~eV) sky. Because
of their large positional uncertainties, these events have not yet been
associated to any astrophysical source. We have found plausible astronomical
counterparts in the GeV -- TeV bands by looking for sources in the available large 
area high-energy $\gamma$-ray catalogues within the
error circles of the IceCube events. We then built the spectral
energy distribution of these sources and compared it with the energy and flux
of the corresponding neutrino. Likely counterparts include mostly
BL Lacs and  two Galactic pulsar wind nebulae. On the one hand many objects,
including the starburst galaxy NGC 253 and Centaurus A, despite being
spatially coincident with neutrino events,  are too weak to be reconciled with 
the neutrino flux. On the other hand, various GeV powerful objects cannot be 
assessed as possible counterparts due to their lack of TeV data. The definitive 
association between high-energy astrophysical neutrinos and our candidates
will be significantly helped by new TeV observations but will be
confirmed or disproved only by further IceCube data. Either
way, this will have momentous implications for blazar jets, high-energy
astrophysics, and cosmic-ray and neutrino astronomy.
 \end{abstract}


\section{Introduction}
\label{sec:Introduction}
The IceCube South Pole Neutrino Observatory\footnote{http://icecube.wisc.edu}
has reported the first evidence of high-energy astrophysical neutrinos\footnote{In this paper 
neutrino means both neutrino and antineutrino.}
\citep{2013PhRvL.111b1103A,ICECube13}, and more recently has
confirmed and strengthened these observations by publishing a sample of 35
events with a deposited energy from 30~TeV to 2~PeV \citep{ICECube14}.
With this enlarged sample the null hypothesis that all events are associated
with the atmospheric background can be rejected at the $5.7\sigma$ level.  If
the observation of ultra-high energy  cosmic rays revealed the
existence of extreme cosmic accelerators, the IceCube neutrinos show that
hadronic particle physics is in action in astrophysical sites at an energy
scale somewhat higher than any man-made accelerator.  IceCube is therefore
opening a new window at the high-energy frontier of particle-
and astro-physics. Motivated by this discovery 
we investigate here plausible 
$\gamma$-ray counterparts of the IceCube events and discuss
possible new scenarios.
 The detection of high-energy neutrinos up to the PeV (10$^{15}$~eV) scale
 implies the existence of a class of astrophysical objects accelerating
 protons up to at least 10$^{16} - 10^{17}$~eV, which then collide with other
 protons ($pp$ collisions) or photons ($p\gamma$ collisions). High-energy
 $\gamma$-rays with energy and flux about a factor two higher than the
 neutrinos at the source, and therefore reaching the $\ga 60$ TeV range for the IceCube events, 
 are also expected
 as secondary products in both cases \citep{2006PhRvD..74c4018K,
 2008PhRvD..78c4013K}. 
 In the following we refer to these $\gamma$-rays as
 neutrino {\it twins}.  The study of these twin photons would provide the
 most direct way to shed light on the origin of the IceCube neutrinos. The
 twin photons, however, cannot be at the moment investigated due to the fact
 that present $\gamma$-ray telescopes reach only $\sim 20 - 40$ TeV.  Moreover,
 depending on the sources and their distance, absorption of
 the twin photons might dilute the direct photon-neutrino connection.
 
The topology of the IceCube detections are broadly classified in two types:
1. cascade-like, characterised by a compact spherical energy deposition;
2. track-like, defined by a dominant linear topology from the induced
muon. 
A large majority of the 35 IceCube events  are characterised by a cascade-like
topology, which, unlike the track-like topology, can only be reconstructed
with a resolution in the tens of degrees.
The association of the IceCube astrophysical neutrinos with
astronomical sources is therefore not only limited by the missing twin
photons but also by the relatively poor resolution of the events.
Moreover, about half of the 35
events are expected to be background produced in the atmosphere (muons and
neutrinos) and so mostly concentrated at lower energies.

IceCube has performed a series of tests on the incoming direction of the
events showing no significant deviation from an assumed isotropic
distribution nor prominent point sources. The all-sky integrated flux of
neutrinos in the 60 TeV -- 3 PeV range is at the level of $10^{-8}$ GeV cm$^{-2}$ s$^{-1}$ sr$^{-1}$ 
per flavour for
an $E^{-2}$ spectrum. We note here that nearly half of the events reported by IceCube are on
the Galactic plane (best fit position $|b_{\rm II}| \le 10^{\circ}$) with a
few relatively close to the Galactic centre. The remaining half are at higher Galactic latitude. 
Hence both Galactic
\citep[e.g.,][]{2013ApJ...774...74F,2014PhRvD..89j3003T} as well as
extra-galactic \citep[e.g.,][]{2014arXiv1403.4089M,2014arXiv1404.6237F}
scenarios provide valid explanations for the IceCube detections.

In this work we study $\gamma$-ray sources which fall within the
median angular error of the IceCube events\footnote{The IceCube Collaboration has reported up to now only the median angular error of their events.}.
We refrain from modelling candidate sources and specific
scenarios, addressing instead the question of the counterparts of the IceCube
events from a purely phenomenological point of view by using the
highest-energy all-sky catalogues. In the absence of photon observations
above 60 TeV, in fact, we investigate possible counterparts in the GeV $-$
TeV band.
Each neutrino is studied independently as if it were a single
detection in the sky. Both Galactic as well as extra-galactic sources are 
treated equally. 

Section 2 defines the list of IceCube events adopted in this paper, while
Section 3 describes the investigated $\gamma$-ray counterparts, their
selection, and their $\gamma$-ray and neutrino {\it hybrid} spectral 
energy distributions (SEDs). Section 4 discusses our results, while in Section 5 we summarise
our conclusions.  We adopt here the definitions used in
\cite{2004vhec.book.....A} for $\gamma$-ray astronomy: ``high energy'' (HE)
or GeV astronomy spans the 30 MeV to 30 GeV energy range while ``very
high energy'' (VHE) or TeV astronomy refers to the 30 GeV to 30
TeV range. For a review of VHE astronomy we refer to \cite{2012APh....39...61H}.


\section{The IceCube Astrophysical Neutrinos}
\label{sec:IceCube}

\begin{table*}
\caption{Selected list of high-energy neutrinos detected by IceCube.}
\begin{tabular}{@{}lllllrrl}
IceCube ID & Dep. Energy & $\nu f_{\nu}$$^a$ & RA (2000) & Dec (2000) & median angular error & $b_{\rm II}$ & Time (MJD)\\
                      & TeV  & $10^{-11}~$erg/cm$^{2}$/s&  &   & deg  &   deg & \\
\hline
#3 & ##78.7$^{+10.8}_{-8.7}$ & $2.0^{+4.5}_{-1.6}$ & 08 31 36 & $-$31 12 00 & $\le$1.4 & $+5$ & 55451.0707415 \\
#4 & #165$^{+20}_{-15}$ & $1.1^{+2.6}_{-0.9}$ & 11 18 00 & $-$51 12 00 & #7.1  & $+9$ & 55477.3930911 \\
#5 & ##71.4$\pm$9.0 & $1.8^{+4.1}_ {-1.5}$ & 07 22 24 & $-$00 24  00 & $\le$1.2  & $+7$ & 55512.5516214 \\
#9 & ##63.2$^{+7.1}_{-8.0}$ & $2.8^{+6.5}_{-2.3}$ & 10 05 12 &  +33 36 00 & 16.5 &  $+54$ & 55685.6629638 \\
10 & ##97.2$^{+10.4}_{-12.4}$ & $1.6^{+3.8}_{-1.4}$ & 00 20 00 & $-$29 24 00& #8.1& $-65$ & 55695.2730442 \\
11 & ##88.4$^{+12.5}_{-10.7}$ & $1.5^{+3.4}_{-1.2}$ & 10 21 12 & $-$08 54 00 & 16.7 & +39 & 55714.5909268 \\
12 & #104$\pm13.0$& $1.2^{+2.8}_{-1.0}$ &19 44 24 & $-$52 48  00 & #9.8 & $-29$ & 55739.4411227\\
13 & #253$^{+26}_{-22}$& $1.6^{+3.7}_{-1.3}$ & 04 31 36 & $+$40 18 00 & $\le$1.2 & $-5$ & 55756.1129755\\
14 & 1041$^{+132}_{-144}$& $1.6^{+3.6}_{-1.3}$ & 17 42 24& $-$27 54  00&13.2 & $-1$ & 55782.5161816\\
17 & #200$\pm$27 & $1.7^{+3.9}_{-1.4}$ & 16 29 36 & $+$14 30  00  & 11.6 & $+38$ & 55800.3755444\\
19 & ##71.5$^{+7.0}_{-7.2}$ & $1.8^{+4.1}_{-1.5}$ & 05 07 36 & $-$59 42 00 & #9.7 & $-36$ & 55925.7958570 \\
20 & 1141$^{+143}_{-133}$& $1.5^{+3.5}_{-1.3}$ & 02 33 12 & $-$67 12 00& 10.7 & $-47$ & 55929.3986232\\
22 & #220$^{+21}_{-24}$& $1.0^{+2.3}_{-0.8}$ & 19 34 48 & $-$22 06 00& 12.1& $-19$ & 55941.9757760\\
26 & #210$^{+29}_{-26}$ & $1.5^{+3.5}_{-1.3}$ & 09 33 36 & $+$22 42 00  &11.8 & $-45$ & 55979.2551738\\
27 & ##60.2$\pm5.6$ & $2.4^{+5.5}_{-2.0}$ & 08 06 48 &  $-$12 36 00 & #6.6 & $+10$ & 56008.6845606 \\
30 & #129$^{+14}_{-12}$& $1.1^{+2.6}_{-0.9}$ & 06 52 48 & $-$82 42 00&  #8.0 & $-27$ & 56115.7283566\\
33 & #385$^{+46}_{-49}$& $1.9^{+4.4}_{-1.6}$ &19 30 00& $+$07 48 00& 13.5 & $-5$ & 56221.3423965\\
35 & 2004$^{+236}_{-262}$& $2.0^{+4.5}_{-1.6}$ & 13 53 36 &$-$55 48 00 & 15.9 & $+6$ & 56265.1338659\\
\hline
\multicolumn{8}{l}{\footnotesize $^a$ Fluxes in units of $10^{-8}$ GeV cm$^{-2}$ s$^{-1}$ can be obtained by
multiplying the numbers in this column by 0.614.}\\
\end{tabular}
\label{tab:ICE}
\end{table*}

The first high-energy astrophysical neutrinos ever detected have been reported
by IceCube through the selection of very high-energy events with
interaction vertex inside the detector \citep{ICECube13}. This containment strategy of very high-energy 
events is extremely efficient in rejecting the atmospheric background, including
atmospheric neutrinos \citep{2009PhRvD..79d3009S}.  In doing so, the analysis
strategy favours cascade-like events, which deposit most of their energy
inside the detector, over track-like ones, which travel various kilometres
outside the detector.  While, on the one hand, through the containment strategy
astrophysical neutrinos are successfully singled out from the entire sky, on the other hand, the
poor resolution of the cascade-like events makes the association with
possible counterparts extremely challenging. Moreover,
the absorption of very high-energy neutrinos crossing the
Earth favours the southern hemisphere over the northern one, contrary to
other IceCube analysis strategies optimized at lower energies, which are more sensitive to northern
hemisphere events \citep{2013ApJ...779..132A}.

In order to reduce the residual atmospheric background contamination, which
might still be produced by mouns and atmospheric neutrinos and
concentrate in the low-energy part of spectrum \citep[see Fig. 2
  in][]{ICECube14}, we consider here only IceCube events with energy $\ge 60$
TeV. Moreover, to somewhat limit the number of possible counterparts,
we consider only events with median angular error $\le
20^{\circ}$. These two cuts reduce the sample from 35 to 18 events.  These
are listed in Tab. \ref{tab:ICE}, which gives the deposited
  energy of the neutrino, the flux at the 
 deposited energy 
 in $\nu f_{\nu}$ units, the coordinates, the median angular error in
degrees, the Galactic latitude, and the time of detection in
Modified Julian Days. By interpreting each single event as coming from one
astrophysical counterpart, in fact, we have derived the flux per neutrino
event assuming that the observed flux is spread over 1 dex in energy and that
the spectrum is $f(\nu) \propto \nu^{-1}$ (equivalent to $dN/dE \propto E^{-2}$).
We use here an energy bin, which is somewhat larger than the IceCube average energy resolution. 
This is done to take into account the larger uncertainty due to the different topologies of the event and possible stochastic variations 
in the deposited energy of the single event. The resulting uncertainty in the flux estimate  
is almost fully absorbed in the Poissonian error and does not affect significantly the comparisons described in the following.
Effective areas from the IceCube 
northern or southern hemisphere (depending on declination) 
and a live time of detection of 988 days \citep{ICECube14} were also used. 
In order to take into account the fact that the deposited energy is $\sim 25\%$ lower than the neutrino energy,
we have averaged the effective areas in two neighbouring bins (at the neutrino
deposited energy and at the next highest one). 
The derived fluxes are in the range $1.0 - 2.8 \times 10^{-11}$ erg
cm$^{-2}$ s$^{-1}$ (i.e., $0.6 - 1.72\times 10^{-8}$ GeV cm$^{-2}$
s$^{-1}$) and errors are Poissonian for one event \citep{1986ApJ...303..336G}. 

The 18 ``golden'' events are assumed here below to be all of astrophysical origin.
They span a range of deposited energies from
 60 TeV to 2 PeV, with a mean of $\sim 350$ TeV. In 
$pp$ or $p\gamma$ collision scenarios \citep{2006PhRvD..74c4018K,
 2008PhRvD..78c4013K}, the energy range of
primary protons would be above a few tens of PeV. In primary cosmic rays,
this is the region around and above the {\it knee}, which can be
interpreted as the cross road of Galactic and extra-galactic cosmic rays
\citep{2014arXiv1403.3380G}. 
Primary cosmic rays up to extremely high energies are significantly
deviated by  magnetic fields, hence the sources of these primary
cosmic rays have not been identified yet.  If the primary cosmic rays
encounter enough target material, they interact producing secondary neutrinos
(from charged mesons) and photons (from neutral mesons).
As photons and neutrinos are neutral they do not feel the effect of  magnetic fields and can therefore 
be used in principle to probe astronomical PeV accelerators, the so-called PeVatrons. 

\section{The counterparts of the IceCube astrophysical neutrinos}
\label{sec:selection}


Neutrinos are intimately connected to their $\gamma$-ray counterparts. 
As a matter of fact, the twin photons coming from the same
interactions that produced the IceCube neutrinos would have energies $\ga 120$ TeV and up to $\sim 4$ PeV, 
and therefore well above the energy range of VHE astronomy. 
We will assume in the following 
a simple direct connection between VHE astronomical data and the $\sim 100$ TeV -- PeV band
sampled by the IceCube data. 
Of course, the larger the gap between the highest photons detected and the IceCube neutrinos 
the more approximate such extrapolation becomes.


\begin{table*}
\caption{$\gamma$-ray--detected blazars in one median angular error radius
  around the positions of the IceCube astrophysical neutrinos.}
\begin{tabular}{@{}llllllrlll}
ID & Catalogue(s) & Counterpart(s)  & RA (2000) & Dec (2000) &  offset & $z$ & f$_{10 - 500 {\rm GeV}}$ & f$_{> 200 {\rm GeV}}$ &Type  \\
                      &                    &       & &  & deg & & $10^{-12}$ c.g.s. & $10^{-12}$ c.g.s.& \\
\hline
#4 & 1FHL & PKS 1101$-$536 & 11 04 15.4 &	$-$53 56 31 & #3.4 & ? & ##4.4 & #... & BL Lac    \\
#9 & TeVCat/WHSP/1FHL &  MKN 421  & 11  04 19.0 & +38 11 41 & 12.8 &  0.031  &  319.8 & $100-500$ & BL Lac   \\   
 & TeVCat/WHSP/1FHL &  1ES 1011+496 & 10 15  04.0 & +49 26 01 & 15.9 &  0.212  &  #38.6 &  ##6.4 & BL Lac  \\   
 & WHSP/1FHL & B2 0912+29           &  09 15 54.0 & +29 32 56 & 11.2& ?  & #17.0 &  #... & BL Lac   \\   
& 1FHL & RX J0908.9+2311&  09 09 21.4 &  +23 12 18 & 16.1&  0.223  & #10.1 & #...& BL Lac     \\   
 & 1FHL & 87GB 105148.6+222705  & 10 54 41.0 &  +22 14  02 & 15.7& ? &  ##5.8 & #... & BL Lac    \\   
 & 1FHL & RX J1100.3+4019  & 11  00 39.6 &  +40 18 54 & 12.9& 0.225   &  ##5.8 & #... & BL Lac \\
& 1FHL & Ton 1015                  &  09 10 40.3 &  +33 33 18 & 11.3& ? &  ##4.9 & #... & BL Lac \\
 & 1FHL & RX J1023.6+3001  & 10 23 38.2 &  +29 59 42 &  #5.3& 0.433  &  ##4.6 &  #... & BL Lac  \\   
 & 1FHL & 1RXS J091211.9+275955  &  09 12 31.7 &  +27 58 26 & 12.6& ?  &  ##4.1 &  #... & BL Lac   \\   
 & 1FHL & B2 1040+24A        & 10 43 17.3 &  +24  07  08 & 12.6& 0.559   &  ##3.1 &  #... & BL Lac  \\   
 & 1FHL & S4 0917+44           &  09 20 55.2 &  +44 43 52 & 14.0& 2.189   &  ##2.5 &  #... & FSRQ \\   
10 &  TeVCat/WHSP/1FHL & H 2356$-$309 & 23 59 09.0 & $-$30 37 22 & #4.7& 0.165 & ##5.6 & ##2.5 & BL Lac    \\
     & 1FHL & RBS 0016 &  00 08 46.6 & $-$23 40 26 & #6.3& 0.147 & ##3.2 & #... & BL Lac    \\     
11 & WHSP &  1RXS J102244.2$-$011257  & 10 22 43.7 & $-$01 13  02 &  #7.7 & 0.369  & ##8.1$^a$ & #... & BL Lac    \\
 & 1FHL & PMN J0953$-$0840          &  09 52 57.1 &  $-$08 39 18 & #7.0&  ?  & #17.6 & #... & BL Lac   \\   
 & 1FHL & 4C +01.28                       & 10 58 30.7 &   +01 33 50 & 14.0& 0.888 & #13.6 & #... & BL Lac      \\   
 & 1FHL & TXS 1013+054                   & 10 16  02.4 &   +05 12 18 & 14.2& 1.714  &  ##6.6 & #... & FSRQ \\ 
 & 1FHL     &  2FGL J1115.0$-$0701 & 11 15  02.4 &  $-$07  01 37 & 13.5& ?   &   ##4.4 &  #... & AGN$^b$  \\  
 & 1FHL & BZB J1107+0222                  & 11  07 30.7 &   +02 23 10 & 16.1& ?  &  ##2.5 & #... & BL Lac   \\   
 & 1FHL & RXS J094620.5+010459  &  09 46 17.8 &  +01 06 18 & 13.3& 0.557  &  ##2.2 &  #... & BL Lac    \\   
 & 1FHL & PKS B1056$-$113               & 10 58 48.7 & $-$11 34 12 &  #9.6& ?   &  ##2.0 & #... & BL Lac     \\      
12 & TeVCat/WHSP/1FHL & PKS 2005$-$489 & 20 09 27.0	& $-$48 49 52 & #5.6 & 0.071 & #39.6 & ##6.7 & BL Lac    \\
     & 1FHL & PMN J1936$-$4719 & 19 36 51.4	 & $-$47 21 22 & #5.6 & 0.265 & #14.5 & #... & BL Lac    \\
13 & 1FHL & 4C +41.11 &  04 23 48.2  & $+$41 51 04 & #2.1$^c$ & ? & #14.8 & #... & BL Lac    \\
14 & 1FHL & VERA J1823$-$3454 & 18 23 43.0 &$-$34 55 08 &  11.3 & ? & #19.4 & #...  & ?    \\
& 1FHL & PMN J1802$-$3940 & 18 02 43.2 &	$-$39 40 26 & 12.5 & 0.296 & ##9.1 & #...  & FSRQ\\
& 1FHL & 1RXS 182853.8$-$241746 & 18 28 59.0	& $-$24 17 02 & 11.1 & ? & ##8.4 & #... & ?\\
17 & TeVCat/WHSP/1FHL & PG 1553+113 & 15 55 44.0 &$+$11 11 41 & #8.9 & ? & 146.7 & ##4.5 & BL Lac   \\
  & 1FHL &  2FGL J1548.3+1453  & 15 48 21.4 &  +14 55 19 & 10.0 & ? & ##4.2 & #... & ?$^d$     \\ 
19 & WHSP/1FHL & 1RXS J054357.3$-$553206           &  05 43 55.4 & $-$55 32 17 &  #6.4& ?  & #14.6 & #... & BL Lac      \\   
 & 1FHL & 1ES 0505$-$546      &  05  06 55.2 & $-$54 34 26 &  #5.1& ? &  ##7.7 &  #... & ?         \\   
 & 1FHL & PKS 0516$-$621                    &  05 16 37.9 & $-$62  09 54 &  #2.7&  1.300  &  ##3.4 & #... & BL Lac       \\   
    & 1FHL & 1RXS J043431.8$-$572718           &  04 34  08.9 & $-$57 25 41 &  #4.9& ?    &  ##1.8 &  #... & ?        \\   
20 & WHSP/1FHL & SUMSS J014347$-$584550 & 01 43 28.8 & $-$58 44 56  & 10.1 & ? & #11.2 & #... & BL Lac   \\ 
    & WHSP & PKS 0352$-$686  & 03 52 57.5 & $-$68 31 17 &  #7.6 & 0.087 & ##... & #... & BL Lac \\
  & 1FHL & 1RXS J024439.8$-$581953  & 02 44 13.9 & $-$58 13 08 & #9.1 & 0.265 & ##6.9 & #... & BL Lac\\ 
22 & WHSP$_{\rm low~bII}$/1FHL & 1H 1914$-$194 & 19 17 45.8 & $-$19 21 54 & #4.8 & 0.137 & #19.2 & #... & BL Lac\\
  & 1FHL & PMN J1921$-$1607 & 19 22 00.2 & $-$16 07 48 & #6.7 & ? & #13.8 & #... & BL Lac\\
 & 1FHL & PKS 1958$-$179 & 20 01 06.7 & $-$17 52 05 & #7.5 & 0.652 & #11.1 & #... & FSRQ\\
 & 1FHL & 1RXS 195815.6$-$301119 & 19 58 24.2	& $-$30 15 18 &  #9.7 & 0.119 & #10.9 & #... &
BL Lac\\
 & 1FHL & PMN J1911$-$1908 & 19 10 55.4 & $-$19 06 14 &  #6.3 & ? & #10.9 & #... & ?\\
26 & WHSP/1FHL & B2 0912+29 & 09 15 54.0 &	$+$29 32 56 & #7.9 & ? & #17.0 & #... & BL Lac    \\
    & WHSP &  2MASXJ09260351+1243341  &  09 26  03.5 & $+$12 43 34 & 10.1 & 0.186 & ##... & #... & BL Lac \\
  & 1FHL & RX J0908.9+2311 & 09 09 21.4 & $+$23 12 18 & #5.6 & ? & #10.1 & #... & BL Lac\\
   & 1FHL & MG1 J090534+1358 & 09 05 39.6	& $+$13 59 10 & 10.9 & ?  & ##9.3 & #... & BL Lac\\
    & 1FHL & OJ 287 & 08 54 50.9 & $+$20 04 44 &  #9.4 & 0.206 & ##4.1 & #... & BL Lac\\
  & 1FHL & 1RXS J091211.9+27595 & 09 12 31.7 & $+$27 58 26 &  #7.1 & ?  & ##4.1 & #... & BL Lac\\
27 & WHSP$_{\rm low~bII}$/1FHL & PMN J0816$-$1311                  &  08 16 21.8 & $-$13 10 37 &  #2.4&  ?  & #18.1 & #... & BL Lac \\
  & 1FHL & PKS 0805$-$07                     &  08  08 16.8 &  $-$07 49 23 &  #4.8&  1.837  & #13.4 &  #... & FSRQ        \\   
  & 1FHL & TXS 0815$-$094                    &  08 18  01.7 &  $-$09 35  06 &  #4.1&  ? &  ##7.3 &  #... & BL Lac \\
   & 1FHL & TXS 0752$-$116                    &  07 54 23.8 & $-$11 49 26 &  #3.1&  ?  &  ##3.7 &  #... & BL Lac \\
 30 & 1FHL & PKS 1029$-$85 & 10 29 00.7 & $-$85 43 48 & #5.9 & ? &  ##3.1 & #... & BL Lac    \\
 & 1FHL & PKS 0736$-$770 & 07 35 05.8	 & $-$77 08 13 & #5.8 & ? &  ##3.0 & #... & BL Lac\\
33 & 1FHL & 1RXS J193109.5+093714 & 19 31 04.8 & $+$09 38 02 & #1.9  & ? & #24.0 & #... & BL Lac \\
 & 1FHL & 1RXS J194246.3+103339 & 19 42 52.1 & $+$10 34 05 & #4.2  & ? & #20.7 & #... & ?\\
35 & TeVCat & 1ES 1312$-$423 & 13 14 58.0 & $-$42 35 49 & 14.6 & 0.105 & #16.5 & ##1.3 & BL Lac \\
     & 1FHL & 1RXS 130737.8$-$425940 & 13 07 43.0 & $-$42 59 56 & 14.8 & ? & #18.8 & #... & ? \\
     & 1FHL & 1RXS 130421.2$-$435308 & 13 04 19.2 & $-$43 54 54 & 14.2 & ? &  #13.0 & #... & BL Lac\\
    & 1FHL & PKS B1424$-$418 & 14 28 02.9 &	 $-$42 06 14 & 14.8 & 1.522 & ##8.9 & #... & FSRQ\\
    & 1FHL & PMN J1234$-$5736 & 12 34 07.4	& $-$57 35 31 & 11.0 & ? & ##7.3 & #... & ?\\
    & 1FHL & PMN J1329$-$5608 & 13 28 46.8	& $-$56 04 44 & #3.5 & ? & ##3.8 & #... & ?\\
    & 1FHL & PMN J1326$-$5256 & 13 27 09.1	& $-$52 58 26 &  #4.8 & ? &  ##3.6 & #... & BL Lac\\ 
    & 1FHL & PKS 1326$-$697 & 13 30 27.6 & $-$70 05 02 & 14.5 & ? & ##2.2 & #... & ?\\
\hline
\multicolumn{7}{l}{\footnotesize $^a$ Flux in the $0.1 - 100$ GeV range}\\
\multicolumn{7}{l}{\footnotesize $^b$ AGN candidate \citep{Ack2012}}\\
\multicolumn{7}{l}{\footnotesize $^c$ Offset $>$ median angular error}\\
\multicolumn{7}{l}{\footnotesize $^d$ Blazar candidate \citep{Mas2013}}\\
\end{tabular}
\label{tab:counterparts}
\end{table*}

\subsection{$\gamma$-ray catalogues}

In order to alleviate the problem of the missing spectral coverage of the $\gamma$-rays twins of the IceCube neutrinos,
our approach is to look for counterparts of the IceCube neutrinos 
in currently available all-sky  catalogues that cover the highest possible energies.
Namely, in decreasing order of energy and priority:

\begin{enumerate}
\item TeVCat\footnote{http://tevcat.uchicago.edu/}, an online catalogue for
  VHE astronomy, which includes at the time of writing 147 sources detected
  typically above $\approx 100$ GeV and reaching the TeV regime. Of these, 53
  are blazars (50 BL Lacs and 3 flat-spectrum radio quasars [FSRQs]; see Sect. 
  \ref{sec:blazar_counterparts} for a description of the two blazar sub-classes) and 59 are 
  Galactic sources, with the rest being mostly unclassified. 
  TeVCat is a list of TeV sources, as there are no all-sky flux-limited TeV catalogues at the 
moment, given the very sparse sky coverage available at these energies. The High Energy 
Stereoscopic System (H.E.S.S.), however, has undertaken a TeV Galactic Plane survey covering 
a range of 250 to 65$^{\circ}$ in longitude at $|b_{\rm II}| < 3.5^{\circ}$ \citep{2013arXiv1307.4868C}. 
There must be, therefore, many more TeV sources with fluxes comparable to 
the detected ones, which are still undetected, particularly at $|b_{\rm II}| > 3.5^{\circ}$; 
\item the Wise High Synchrotron Peaked (WHSP) catalogue \citep{arsioli2014}, which 
provides a large area ($|b_{\rm II}| > 20^{\circ}$) catalogue of $\sim 1,000$ blazars and blazar candidates selected 
to have the peak of the synchrotron emission at $\nu > 10^{15}$ Hz and therefore expected 
to radiate strongly in the HE and VHE bands. For our purposes we selected the sub-sample 
of 76 sources with a ``figure of merit'' (FoM) on their potential detectability in the TeV band $\ge 1.2$. 
The FoM is defined as the ratio between the synchrotron peak flux of a source and that of the 
faintest blazar in the WHSP sample already detected in the TeV band.
The WHSP sources are all BL Lacs,  
with $39\%$ of them being known TeV sources and the remaining ones thought to be within
reach of detection by current VHE instrumentation. Although technically not a $\gamma$-ray
catalogue, this WHSP sub-sample represents at present the best way to compensate for 
the lack of full sky coverage in the TeV band for blazars. Moreover, $\sim 91\%$ of these sources
have a {\it Fermi~} 2FGL or 3FGL $\gamma$-ray counterpart \citep{arsioli2014}; 
 \item the first {\it Fermi}-Large Area Telescope (LAT) catalogue of sources
  detected above 10 GeV (1FHL), which includes 514 sources, $\sim 75\%$ of
  which are blazars (277 BL Lacs and 53 FSRQs) or blazar candidates (58)
  \citep{1FHL}. The remaining objects are unclassified ($\sim 13\%$) and
  Galactic ($\sim 11\%$).
\end{enumerate} 

Our logic can be thus summarized: in the absence of VHE data reaching
the $\sim 100$ TeV -- PeV band, we use TeVCat as our starting point. To
compensate for its incompleteness and limited sky coverage we then add the
WHSP catalogue, which however covers the high Galactic latitude sky and includes only
blazars. Furthermore, as WHSP provides for most of its sources only a FoM for
the TeV detectability, we complement it by using the highest energy {\it
  Fermi} catalogue, that is 1FHL, which gives an all-sky view above 10 GeV
for all astronomical sources. As it turns out, the gap between 10 GeV and the
neutrino energies ($\ge 60$ TeV) appears to be quite large for a sensible
extrapolation. Nevertheless, we wanted to be as thourough as possible in our
search without penalizing sources, which have not been observed yet in the
TeV band but might still one day prove to be plausible IceCube counterparts.

\subsection{Blazar counterparts}\label{sec:blazar_counterparts}

Blazars are those Active Galactic Nuclei (AGN) whose emission is dominated by a relativistic jet 
viewed at a relatively small angle with respect to the line of sight \citep{1995PASP..107..803U}. 
The two main blazar sub-classes, namely BL Lacertae objects (BL Lacs) and FSRQs, differ mainly in
their optical spectra, with the former displaying strong, broad emission lines and the latter instead being 
characterized by optical spectra showing at most weak emission lines, sometimes exhibiting absorption 
features, and in many cases being completely featureless \citep[see][for a recent re-evaluation of the two 
blazar classes]{2012MNRAS.420.2899G,2013MNRAS.431.1914G}. The strong
non-thermal blazar radiation, which spans the entire electromagnetic spectrum, is composed of two 
broad humps, the low-energy one attributed to synchrotron radiation, and the high-energy one,
usually thought to be due to inverse Compton radiation \citep[see e.g.][]{abdosed} or, alternatively, to 
hadronic processes \citep[e.g.][and reference therein]{2013ApJ...768...54B}. The peak of 
the synchrotron hump, $\nu_{\rm peak}$, ranges from about $\sim 10^{12.5}$~Hz to over
$10^{18}$~Hz reflecting the maximum energy at which particles can be
accelerated \citep[e.g.][]{GiommiPlanck}. Blazars with $\nu_{\rm peak} < 10^{14}$~Hz 
in their rest frame are called Low Synchrotron Peaked (LSP)
sources, while those with $10^{14}$~Hz $<  \nu_{\rm peak} < 10^{15}$~Hz, and
$\nu_{\rm peak} > 10^{15}$~Hz are called Intermediate and High Synchrotron Peaked (ISP and
HSP) sources respectively \citep{abdosed}. This definition extends the
original division of BL Lacs into LBL and HBL sources first introduced by
\cite{padgio95}. Basically all HSP are BL Lacs \citep[see][for a possible explanation]
{2012MNRAS.420.2899G,2012MNRAS.422L..48P}.
Objects with large $\nu_{\rm peak}$ are obviously favoured to be
VHE sources. Indeed, only one out of the fifty currently TeV-detected BL Lacs is an LSP. 

Due to their very large luminosities and SEDs routinely reaching the HE and VHE bands, blazars are thought to be amongst 
the most powerful accelerators in the Universe and as a result have been considered 
prime candidate sources of ultra-high energy cosmic rays and neutrinos
\citep{1993ICRC....1..447H, 1997ASPC..121..585P, 1999APh....11...49M, 2001astro.ph..7200D}.

Tab. \ref{tab:counterparts} shows the results of our search for blazars and
gives the IceCube ID, the catalogues where the counterparts were found, the
counterparts' names and coordinates, ranked by energy (TeVCat first) and
flux, the offset between the reconstructed position of the IceCube event and the blazar
 one, the redshift of the source (if available), the $10 - 500$ GeV flux (from the 1FHL
catalogue), the (observed) flux above 200 GeV (from papers referenced in TeVCat) for the
TeV-detected sources, and the blazar type, namely BL Lac, FSRQ, or
unknown\footnote{This refers to the so-called ``active galaxies of uncertain
  type'' (AGU), most of which are expected to be blazars
  \citep[e.g.][]{1FHL}.}. Note that given the very strong variability of
blazars flux values should be taken only as approximate. Nevertheless,
fluxes are important as, on average, a stronger neutrino source should
also be a stronger $\gamma$-ray source, unless significant absorption is
present (Sect. \ref{sec:Introduction}). 

Blazar counterparts were found for 16/18 neutrino events, in one case (ID 13)
with an offset slightly larger ($\sim 1.8$ times) than the median angular error. 
Namely, six (with two sources correlated with the
same event) were found in TeVCat (all of them in 1FHL), eleven 
are WHSP sources (nine of which are in 1FHL as well), while all
others were found in 1FHL. 

We stress that 8/9 of the IceCube events with $|b_{\rm II}| > 20^{\circ}$ have WHSP counterparts. 
Even more strikingly, and as is the case for TeVCat, in all these cases the WHSP source(s) 
is (are) always the strongest one(s). This vindicates the use of a selection on synchrotron peak and
flux to identify present or potential TeV emitters and is particularly important for events without TeVCat
counterparts, namely ID 11, 19, 20, and 26. We also note that the strongest 1FHL counterpart of ID 22,
1H 1914$-$194, which has $|b_{\rm II}| < 20^{\circ}$ and therefore is not included by definition in WHSP, 
fullfills the other criteria ($\nu_{\rm peak}$ and FoM) and therefore is a very promising TeV candidate. Same 
story for PMNJ0816$-$1311, the strongest 1FHL counterpart of ID 27\footnote{We thank Paolo Giommi for pointing
this out to us.}. We have therefore marked these two sources as WHSP$_{\rm low~bII}$ in Tab. \ref{tab:counterparts}.

Tab. \ref{tab:counterparts}
includes some well-known, bright BL Lacs, i.e., MKN 421, PKS 2005$-$489, PG
1553+113. Only three sources, namely B2 0912+29, RX J0908.9+2311, and 1RXS
J091211.9+275955, are associated with more than one
neutrino event (ID 9 and 26). Out of the 61 unique objects in
Tab. \ref{tab:counterparts} only six are FSRQs, with none of them being the
strongest source within the error circle. Most ($\sim 88\%$) neutrino events
have more than one blazar counterpart. 

For the two neutrino events for which no counterpart was found in the three
catalogues used, we checked for completeness the {\it Fermi}-2FGL catalogue
  \citep{2FGL}, which includes 1,873 sources detected above 100 MeV\footnote{The 2FGL
  includes blazars or blazar candidates ($\sim 59\%$), Galactic ($\sim 10\%$) and 
  unclassified ($\sim 31\%$) sources.}. In both cases we found a counterpart: 
  2FGL J0825.9$-$3216 (PKS 0823$-$321, an AGU), with an offset of 1.6$^{\circ}$ (slightly above the 
  median angular error) for ID 3 and 2FGL J0726.0$-$0053 (PKS 0723$-$008, an other AGU), 
   with an offset of 1$^{\circ}$ for ID 5. For both objects the HE emission reaches $\sim 6$ GeV, i.e. not 
   too far from the 1FHL cutoff.
 
\begin{figure}
\includegraphics[height=9cm]{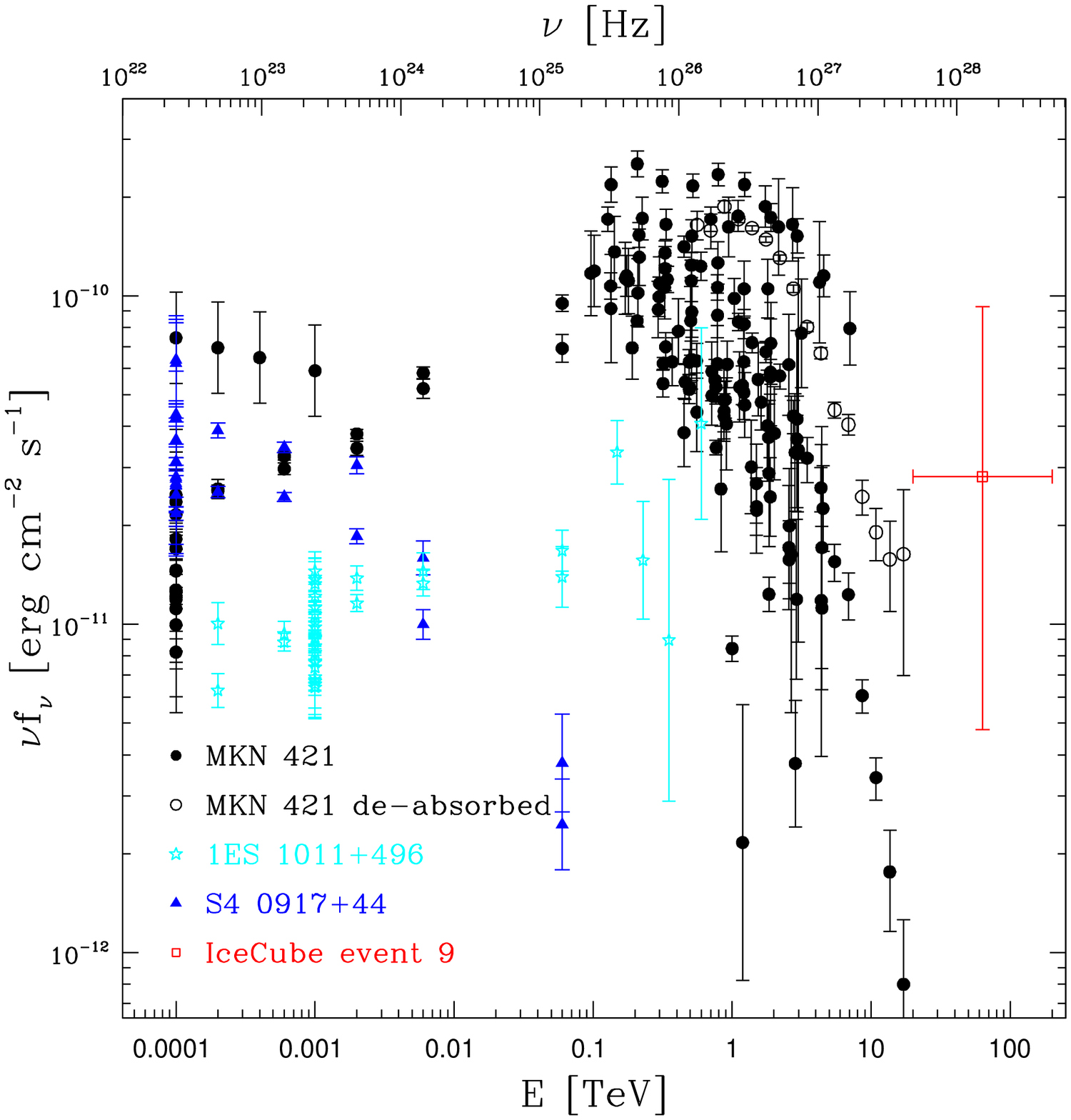}
\caption{$\gamma$-ray SED of MKN 421 (black filled circles) and the HEGRA data \citep{2002A&A...393...89A} 
corrected for absorption by the EBL 
(black open circles) \citep{2008A&A...487..837F}, 1ES 1011+496 corrected for absorption 
by the EBL (cyan stars), and S4 0917+44 (blue triangles), respectively
the strongest, second strongest, and weakest sources in the IceCube error circle (ID 9). The many data
points for MKN 421 represent different states of the source. 
The (red) open square represents the neutrino flux for the corresponding IceCube event; 
vertical error bars are Poissonian for one event, while the horizontal one indicates the range over which
the flux is integrated.}
\label{SED_MKN421}
\end{figure}

\begin{figure}
\includegraphics[height=9cm]{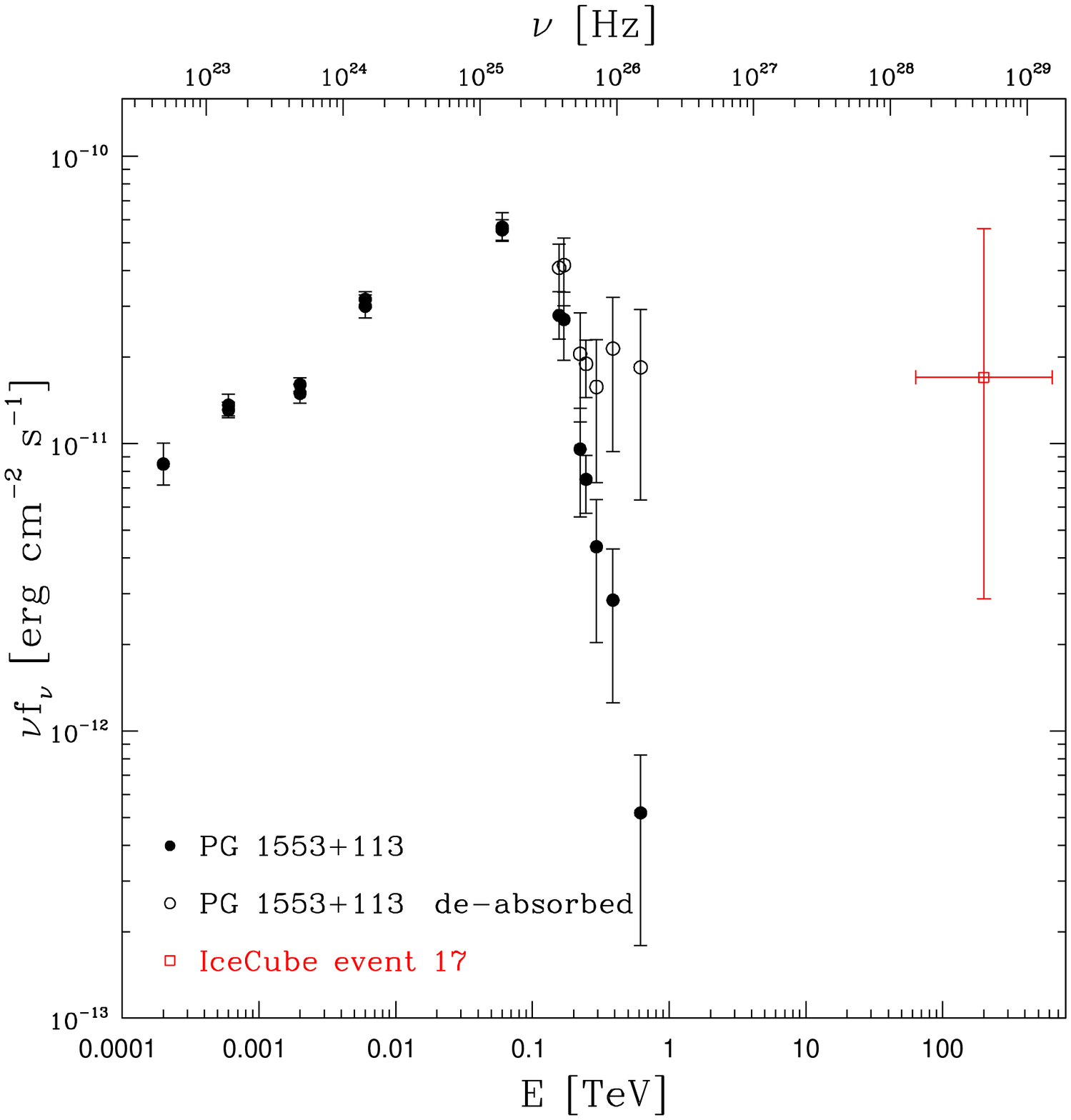}
\caption{$\gamma$-ray SED  of PG 1553+113: observed (black filled circles) 
and corrected for absorption by the EBL (black open circles) assuming 
$z=0.4$ \citep{2012ApJ...748...46A}. The (red) open 
square represents the neutrino flux for the corresponding IceCube event (ID 17); error bars as
described in Fig. \ref{SED_MKN421}.}
\label{SED_1553}
\end{figure}

\begin{figure}
\includegraphics[height=9cm]{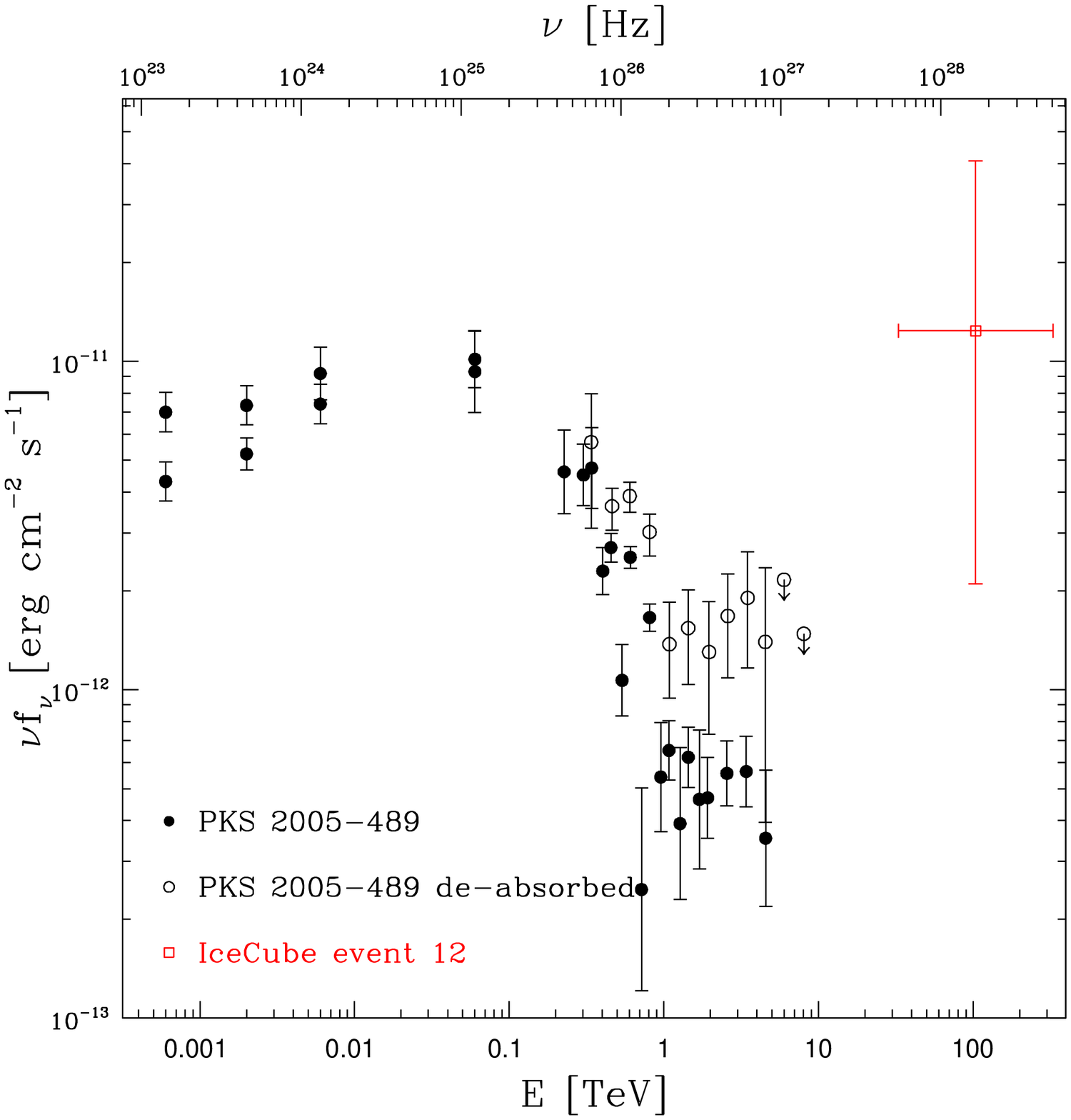}
\caption{$\gamma$-ray SED  of PKS 2005$-$489: observed (black filled circles) 
and corrected for absorption by the EBL (black open circles) \citep{2013ApJ...764..119S}. 
The (red) open square represents the neutrino flux for the corresponding IceCube event (ID 12); error bars as
described in Fig. \ref{SED_MKN421}.}
\label{SED_2005}
\end{figure}

\begin{figure}
\includegraphics[height=9cm]{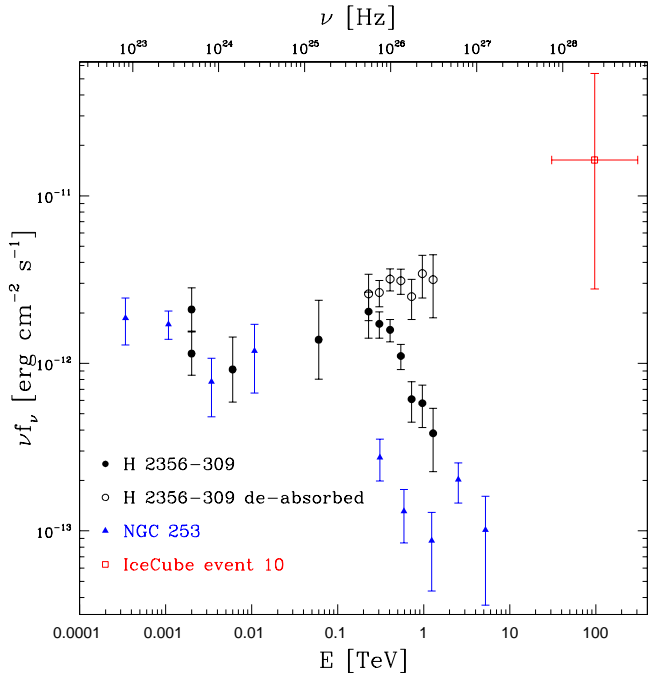}
\caption{$\gamma$-ray SED  of H 2356$-$309: observed (black filled circles) 
and corrected for absorption by the EBL (black open circles) \citep{Costa12}. The SED of NGC 253 (blue triangles) is also
shown \citep{2012ApJ...757..158A}. The (red) open square represents the neutrino flux for the corresponding
  IceCube event (ID 10); error bars as
described in Fig. \ref{SED_MKN421}.}
\label{SED_2356}
\end{figure}

\subsubsection{Hybrid SEDs}\label{sec:hybrid_bl}

To see how the neutrino and photon energetics compare, we have put together
the $\gamma$-ray SEDs of all sources using the SED
builder\footnote{http://tools.asdc.asi.it/SED/} of the ASI Science Data
Centre (ASDC) adding, if needed, VHE data 
taken from the literature. We have also included the flux per neutrino
event at the specific energy, thereby building a hybrid photon -- neutrino SED. 
We then 
performed an ``energetic'' diagnostic by checking if a simple extrapolation succeeded
in connecting the most energetic $\gamma$-rays to the IceCube neutrino in the hybrid SED,
taking into account the rather large uncertainty in the flux of the latter.
If this was the case we considered the source to be a probable counterpart.
Otherwise, we discarded the object. Anything more sophisticated would require detailed modelling,
which goes beyond the scope of this paper. 
We show in Figs. \ref{SED_MKN421} -- \ref{SED_2356} and Fig. \ref{SED_ID35} the SEDs of 
the TeV-detected blazars, which we now turn to comment:

\begin{itemize}

\item MKN 421\footnote{\cite{2014arXiv1404.6237F} have compared 
 the TeV flux  of MKN 421 and the neutrino fluxes related to ID 9.} (Fig. \ref{SED_MKN421}); this is the strongest $\gamma$-ray 
  source in our sample and a 
simple extrapolation of its VHE spectrum has no problems in explaining the energy and flux of the corresponding
IceCube event (ID 9), even taking into account the factor $\sim 2$ increase expected for the energy and flux 
of the twin photons associated with the neutrinos. This is a clear case that passes the  ``energetic'' diagnostic. 
Fig. \ref{SED_MKN421} shows also the 
SED, corrected for absorption by the extragalactic background light (EBL), of 1ES 1011+496, the other 
TeVCat BL Lac in the same error circle, which appears also to be a plausible counterpart. Finally, the SED of S4 
0917+44, which is the weakest source amongst the other nine blazars associated with ID 9, is also displayed. 
While the $\gamma$-ray spectra
of MKN 421 and 1ES 1011+496 are raising with energy, that of S4 0917+44 is falling, which makes 
it a very unlikely counterpart for the IceCube event. 
This shows how important the ``energetic'' diagnostic 
is, especially because of the relatively loose spatial one, given the large error circles. 
We note that, for our purposes, the extrapolation 
to $\gamma$-ray energies larger than the observed ones 
is not influenced by the EBL as neutrinos reflect the photon densities {\it at the source} and not at the
observer after the interaction with the EBL photons. However, the comparison
between observed photon and neutrino fluxes is affected, as the former will
be absorbed by the EBL, above $\approx 100$ GeV, while the latter will not;

\item PG 1553+113 (Fig. \ref{SED_1553}); in this case the EBL has a very strong effect  
and, after correcting for its absorption \citep[assuming $z=0.4$: see discussion in][]{2012ApJ...748...46A}, 
its VHE spectrum appears also consistent with the IceCube event (ID 17);

\item PKS 2005$-$489 (Fig. \ref{SED_2005}); despite being relatively local, the EBL has an effect also on the
VHE spectrum of this BL Lac, which reaches relatively large energies ($\sim 5$ TeV). However, even taking
this into account, its VHE spectrum appears to be inconsistent with a neutrino detection (ID 12);

\item H 2356$-$309 (Fig. \ref{SED_2356}); the EBL has quite a strong effect also on the VHE spectrum of this BL 
Lac and, after correcting for it, its VHE spectrum appears consistent with the IceCube event 
(ID 10). 

\end{itemize}

\begin{figure}
\includegraphics[height=9cm]{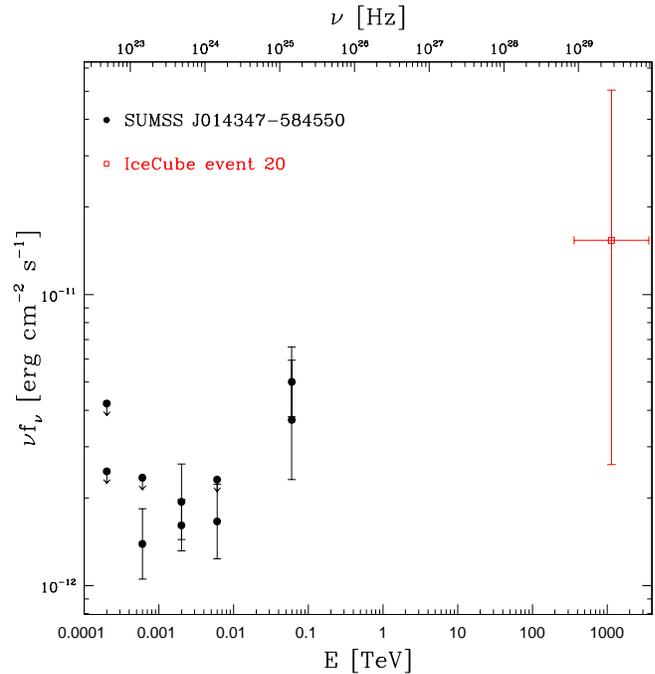}
\caption{$\gamma$-ray SED  of SUMSS J014347$-$584550 (black filled circles). The (red) open square 
represents the neutrino 
flux for the corresponding IceCube event (ID 20); error bars as
described in Fig. \ref{SED_MKN421}.}
\label{SED_SUMSS_0143}
\end{figure}

We point out that also some of the 1FHL blazars in Tab. \ref{tab:counterparts} display SEDs, which appear
at a first glance not inconsistent with the corresponding neutrino event. However, 
the gap between the highest observable energy and that of the neutrino is much ($\sim 10 - 100$ times) 
larger than that typical of TeV-detected sources, making any possible neutrino -- photon association 
harder to pin down. We make an exception for WHSP sources, for which we have very strong
hints of a possible TeV emission. As an example, we show in Fig. \ref{SED_SUMSS_0143} the SED of 
SUMSS J014347$-$584550, a WHSP source and the strongest counterpart of ID 20 (one of the three
PeV events). A simple extrapolation of the HE spectrum appears not inconsistent with the energy and flux 
of the corresponding IceCube event, given its rising SED. The same argument applies to 
1RXS J054357.3$-$553206 (ID 19), 1H 1914$-$194 (ID 22) and PMN J0816$-$1311 (ID 27). 

We present our list of the most plausible counterparts in Sect. 
\ref{sec:disc}. 

\begin{table*}
\caption{$\gamma$-ray--detected non-blazars in one median angular error
  radius around the positions of the IceCube astrophysical neutrinos.}
\begin{tabular}{@{}lllllllll}
ID & Catalogue & Counterpart(s)  & RA (2000) & Dec (2000) &  offset & f$_{10 - 500 {\rm GeV}}$ & flux  & Class$^a$  \\
                      &                    &       & &  &  deg & $10^{-12}$ c.g.s. &Crab & \\
\hline
 10 & TeVCat &       NGC 253	                  &  00 47 34 & $-$25 17 22 &  #7.4 & #0.6$^b$ & 0.0021 & {\bf Starburst}   \\  
 14$^c$   & TeVCat &       HESS J1804$-$216             & 18  04 31 & $-$21 41 60 &  #8.0  & #... & 0.25 & extended	       \\  
    & TeVCat &       HESS J1809$-$193             & 18 10 31 & $-$19 18  00 & 10.7  & #... & 0.14 & PWN         \\  
    & TeVCat &       HESS J1813$-$178             & 18 13 36 & $-$17 50 24 & 12.4 & #... & 0.06 & PWN         \\  
    & TeVCat &       HESS J1745$-$303             & 17 45  02& $-$30 22 12 &  #2.5 & #... & 0.05  & SNR/Molec   \\  
    & TeVCat/1FHL &       Galactic Centre              & 17 45 39 & $-$29  00 22 &  #1.3  & 30.2 & 0.05 & unidentified        \\  

    & TeVCat &       CTB 37A	                  & 17 14 19 & $-$38 34  00 & 12.2 & #... & 0.03& SNR/Molec   \\  
    & TeVCat &       HESS J1718$-$385             & 17 18  07 & $-$38 33  00& 11.8 & #... & 0.02& PWN	        \\  
    & TeVCat &       HESS J1741$-$302             & 17 41  00& $-$30 12  00&  #2.3 & #... & 0.01 & unidentified	       \\  
    & TeVCat &       G0.9+0.1                     & 17 47 23 & $-$28  09  06 &  #1.1 & #... & 0.02 &  PWN         \\  
    & TeVCat &       CTB 37B                      & 17 13 57 & $-$38 12  00 & 11.9 & #... &0.018 & Shell	      \\  
    & TeVCat &       Terzan 5                     & 17 47 49 & $-$24 48 30 &  #3.3  & #... & 0.015 & Globular    \\  
   & TeVCat/1FHL &       SNR G349.7+0.2               & 17 18  01& $-$37 26 30 & 10.8  &  19.9 &0.004 & SNR/Molec   \\  
    & TeVCat &       HESS J1729$-$345             & 17 29 35 & $-$34 32 22 &  #7.2 & #... & #...& extended       \\  
    & TeVCat &       HESS J1731$-$347             & 17 32  03 & $-$34 45 18 &  #7.2 & #... &#... & Shell	      \\  
    & TeVCat &       HESS J1800$-$240C            & 17 58 51 & $-$24  03 07 &  #5.3  & #... & #... & SNR/Molec   \\  
    & TeVCat &       HESS J1800$-$240B            & 18  00 26 & $-$24  02 20 &  #5.6 & #... &  #... & SNR/Molec   \\  
    & TeVCat/1FHL &       W28                          & 18  01 42 & $-$23 20  06 &  #6.3  & 46.3 & #... & SNR/Molec   \\  
    & TeVCat &       HESS J1800$-$240A            & 18  01 57 & $-$23 57 43 &  #5.9  & #... & #... & SNR/Molec   \\  
    & TeVCat &       HESS J1808$-$204              & 18  08  00 & $-$20 24  00 &  #9.5  & #... &#...  & unidentified	       \\  
    & 1FHL & PWN G0.13$-$0.11                     & 17 46 22 & $-$28 51 47 &  #1.3 & 30.2 &   & PWN         \\ 
    & 1FHL &                                      & 17 58 22 & $-$23 40 16 &  #5.5 & 12.9 &  &  unidentified              \\ 
    & 1FHL & LAT PSR J1809$-$2332                 & 18  09 51 & $-$23 29 46 & #7.6 & 10.1 &   & HPSR        \\ 
    & 1FHL &                                      & 17 41 56 & $-$25 39 29 &  #2.2 & #7.7 &  & unidentified               \\ 
19 & 1FHL & Large Magellanic Cloud   &  05 26 36 & $-$68 25 12 &  #9.0&  38.8 & &  {\bf Galaxy}         \\  
  & 1FHL &                                    &  05  09 56 & $-$64 19 44 &  #4.6&  #3.5 &  & unidentified             \\  
 27 & 1FHL &                                 &  08  04 53 &  $-$06 26  06 &  #6.2&  #2.3 &   &  unidentified       \\  
 33  & TeVCat/1FHL &       MGRO J1908+06                & 19 07 54 & +06 16  07 &  #5.7  & #5.1 & 0.17 & PWN$^d$      \\  
    & TeVCat &       HESS J1912+101               & 19 12 49 & +10  09 06 &  #4.9 & #... &0.1  & PWN	        \\  
      & TeVCat/1FHL &       W51                          & 19 22 55 & +14 11 27 &  #6.6  & 55.2 & 0.03 & SNR/Molec   \\  
    & TeVCat &       G54.1+0.3                    & 19 30 32 & +18 52 12 & 11.1  & #... & 0.025 & PWN	        \\  
     & TeVCat &       IGR J18490$-$0000            & 18 49  01& $-$00  01 17 & 12.9  & #...  &0.015 &  PWN	        \\  
               & TeVCat/1FHL &       W49B                         & 19 11  06 & +09  05 34 &  #4.8 & 27.7 & 0.005 & SNR/Molec   \\  
    & TeVCat/1FHL &       HESS J1857+026               & 18 57 11 & +02 40  00 &  #9.6 & 64.2 & #... & extended       \\  
    & TeVCat &       HESS J1858+020               & 18 58 20 & +02 05 24 &  #9.7  & #... & #...  & extended       \\  

%
    & 1FHL & SNR G034.7$-$00.4                    & 18 55 58 &   +01 21 18 & 10.6 & 27.3 &     & SNR         \\ 
 35 & TeVCat &       HESS J1303$-$631             & 13  02 48 & $-$63 10 39 &  #9.8  & #... &0.17 &  PWN         \\  
    & TeVCat &       MSH 15$-$52                  & 15 14  07 & $-$59  09 27 & 11.3  & #... & 0.15& PWN         \\  
        & TeVCat &       HESS J1356$-$645             & 13 56  00 & $-$64 30  00 &  #8.7  & #... & 0.11& PWN         \\  
    & TeVCat &       RCW 86                       & 14 42 42 & $-$62 26 41 &  #9.1  & #... & 0.1 & Shell       \\  

    & TeVCat &       PSR B1259$-$63               & 13  02 49 & $-$63 49 53 & 10.2 & #... &0.07  & Binary      \\  
    & TeVCat/1FHL &       Kookaburra J1420$-$607             & 14 20  09 & $-$60 45 36 &  #6.1  & 31.8 (23.4$^b$)  & 0.07& PWN         \\  
    & TeVCat/1FHL &       Kookaburra J1418$-$609          & 14 18  04 & $-$60 58 31 &  #6.1  & 12.1  (17.4$^b$) & 0.06& PWN         \\  
    & TeVCat &       HESS J1458$-$608             & 14 58  09 & $-$60 52 38 &  #9.8  & #... &0.06 &  PWN         \\  
    & TeVCat &       HESS J1503$-$582             & 15  03 38 & $-$58 13 45 &  #9.8 & #... & 0.06 & DARK        \\  
    & TeVCat/1FHL &       HESS J1507$-$622             & 15  06 52 & $-$62 21 00 & 11.4  & #9.0& 0.01& extended	\\  
    & TeVCat/1FHL &       Centaurus A                  & 13 25 26 & $-$43  00 42 & 13.6  & #6.1 #(1.7$^b$) & 0.008&  {\bf Radio galaxy}        \\  
          & TeVCat &       HESS J1427$-$608             & 14 27 52 & $-$60 51  00 &  #6.8  & #... &  #...& extended 	\\  
    & TeVCat &       G318.2+0.1                   & 14 57 46 & $-$59 28  00 &  #9.3  & #... & #... & SNR/Molec   \\  
    & 1FHL &                                      & 14  07 12& $-$61 33 40 &  #6.0 & 21.3 &  & unidentified            \\ 
    & 1FHL &                                      & 13 28 35   & $-$47 28 16 &  #9.2 & #9.1 & & unidentified           \\ 
    & 1FHL & LAT PSR J1413$-$6205                 & 14 13 25 & $-$62  05 53 &  #6.8 & #8.8 &      & HPSR        \\ 
    & 1FHL &                                      & 13 53  05 & $-$66 42 43 & 10.9 & #6.8 & & unidentified          \\ 
    & 1FHL & PSR J1514$-$4946                     & 15 14 20& $-$49 45  04 & 13.6 & #3.2 &  &   HPSR        \\ 
\hline
\multicolumn{9}{l}{\footnotesize $^a$ Binary: $\gamma$-ray binary; DARK: no
  associations in other bands; extended: unclassified Galactic source; Globular: globular cluster;}\\
\multicolumn{9}{l}{\footnotesize HPSR: Pulsar
  identified by pulsations above 10 GeV; PWN: pulsar wind nebula; Shell:
  shell-type supernova remnant; SNR: supernova remnant;}\\
  \multicolumn{9}{l}{\footnotesize SNR/Molec: supernova
  remnant/molecular cloud}\\
\multicolumn{9}{l}{\footnotesize $^b$ Flux at energies ${> 200~{\rm GeV}}$}\\
\multicolumn{9}{l}{\footnotesize $^c$ Event consistent with the position of
  the Galactic centre}\\
\multicolumn{9}{l}{\footnotesize $^d$ \cite{2013ApJ...773...77A}}\\
\end{tabular}
\label{tab:non_counterparts}
\end{table*}

\subsection{Non-blazar counterparts}\label{sec:non-blazar}
Alternative scenarios for PeV neutrino sources beyond the blazar one include $\gamma$-ray 
bursts (GRB) \citep[and references therein]{2014ApJ...785...54A}, supernovae remnants \citep[and references 
therein]{2008PhRvD..78j3007V}, pulsar wind nebulae \citep[PWN;][]{2003A&A...407....1B}, and 
star-forming galaxies  \citep[and references therein]{2014arXiv1404.1189T}.
At present, the IceCube Collaboration reports no evidence of a GRB -- neutrino connection 
\citep{2012Natur.484..351A}, although this not based on the same events considered here. 

The non-blazar counterparts present in the catalogues studied in this work are listed in
Tab. \ref{tab:non_counterparts}, where we give also the $10 - 500$ GeV flux
(from the 1FHL catalogue) and the HE flux in Crab units (as provided by TeVCat), 
if available. Sources were found for six neutrino
events, four of them on the Galactic plane ($|b_{\rm II}| \le 10^{\circ}$)
and one of which actually consistent with the position of the Galactic
centre. The error circles of the two events off the Galactic plane
include one and two non-blazar counterparts respectively: NGC 253, a
starburst galaxy (ID 10), and the Large Magellanic Cloud (LMC) plus an
unidentified source (ID 19). While NGC 253 and the unidentified source have
fluxes $\sim 4$ times smaller than those of the brightest blazars in
the same error circle, this is not the case for the LMC, which is $\sim 2.6$
times brighter than 1RXS J054357.3$-$553206, which however is a WHSP source. 
ID 27 has a single, weak non-blazar counterpart which, being
an unidentified 1FHL source, might be a blazar as well. For the other three events we found a
very large number ($\sim 10 - 20$) of counterparts, many with fluxes larger
than those of the blazar counterparts. For example, Kookaburra J1420$-$607,  
a PWN 
associated with ID 35 has a flux above 200 GeV $\sim 20$ times larger than
that of 1ES 1312$-$423, the BL Lac in the same error circle in TeVCat. And two
supernova remnants associated with ID 14 and 33 have a $10 - 500$ GeV flux
$\sim 2.5$ times larger than the respective brightest associated BL Lacs. We address
in more detail the comparison between blazar and non-blazar counterparts in 
Sect. \ref{sec:hybrid_non_bl}.

We note that of the thirty-seven classified sources in
Tab. \ref{tab:non_counterparts} all but three are Galactic, the exceptions
being (highlighted in bold face in the table) the LMC, NGC 253, mentioned above, and Centaurus A, a radio
galaxy associated with ID 35. This latter source is particularly interesting
in this respect as the region around it is populated by a number of ultra
high-energy cosmic ray events larger than the rest of the sky, although
recent results show a chance probability for this to occur at a level of
$\sim 4\%$ \citep{Kam12}. However, we point out that, amongst the many
$\gamma$-ray sources in the error circle of ID 35, Centaurus A is one of the
weakest.

\subsubsection{Hybrid SEDs}\label{sec:hybrid_non_bl}

As done for blazars, we have put together the hybrid SEDs of all sources. It turned out that 
only two Galactic sources passed the ``energetic'' diagnostic being actually better at that
than the blazars in the same error circle. We show here some of 
the most promising Galactic counterparts of the IceCube events. 

\begin{figure}
\includegraphics[height=9cm]{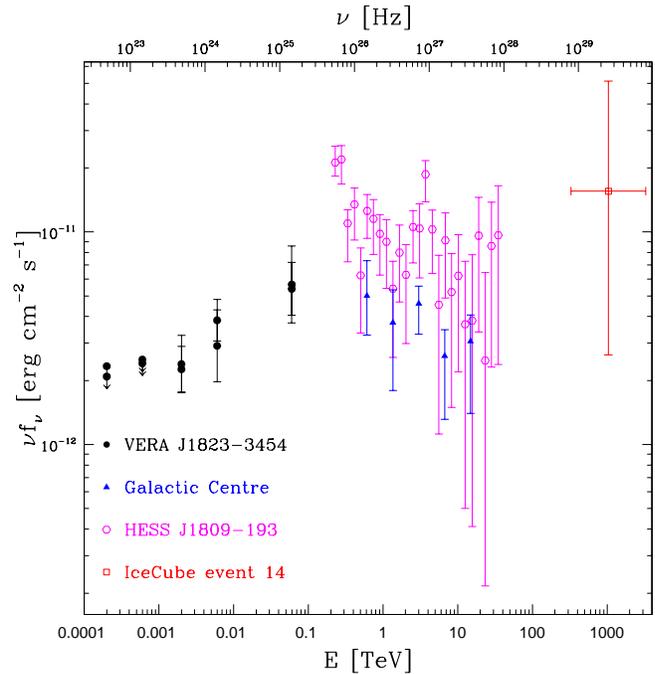}
\caption{$\gamma$-ray SEDs of three sources in the error circle of ID 14, namely: 
the BL Lac VERA J1823$-$3454 (black filled circles), 
the PWN HESS J1809$-$193 (magenta hexagons) \citep{2008AIPC.1085..285R} and
the Galactic Centre \citep{2004A&A...425L..13A}. 
The (red) open square represents the neutrino flux for the corresponding IceCube event; 
error bars as described in Fig. \ref{SED_MKN421}.}
\label{SED_ID14}
\end{figure}

Fig. \ref{SED_ID14} shows the $\gamma$-ray SEDs of three sources in the error circle of ID 14,
namely VERA J1823$-$3454, the strongest blazar, HESS J1809$-$193, 
the second strongest TeV source and a PWN, and the Galactic Centre. 
The latter two reach $\sim 20 - 40$ TeV but only HESS J1809$-$193 seems to be 
a plausible astronomical counterpart, since the Galactic Centre appears to be too soft. 
The BL Lac has a strongly rising SED but reaches only $\sim 60$ GeV, which makes any 
extrapolation to the PeV range very uncertain. 

\begin{figure}
\includegraphics[height=9cm]{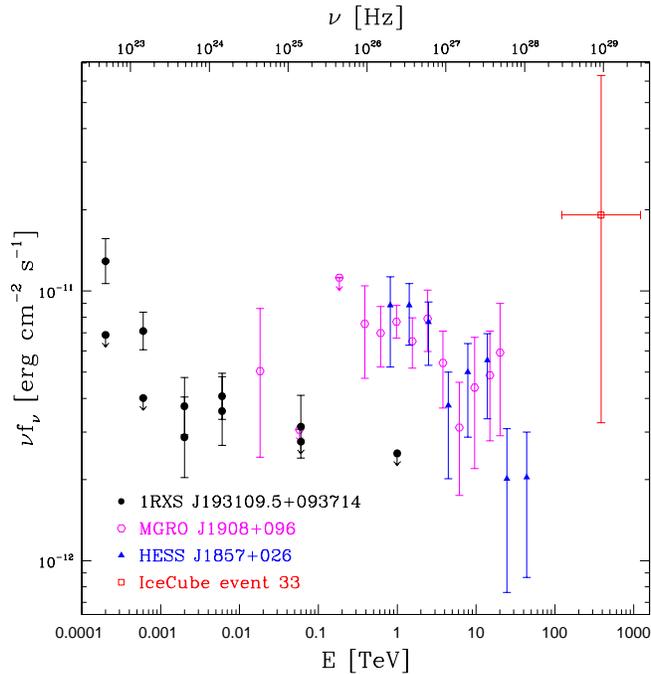}
\caption{$\gamma$-ray SEDs of three sources in the error circle of ID 33, namely: 
the BL Lac 1RXS J193109.5+093714 (black filled circles) and the Galactic
sources MGRO J1908+06 (magenta hexagons) \citep[][and references therein]{2013ApJ...773...77A}
and HESS J1857+026 (blue triangles) \citep{2008A&A...477..353A}. 
The (red) open square represents the neutrino flux for the corresponding IceCube event; error bars as
described in Fig. \ref{SED_MKN421}.}
\label{SED_ID33}
\end{figure}

\begin{figure}
\includegraphics[height=9cm]{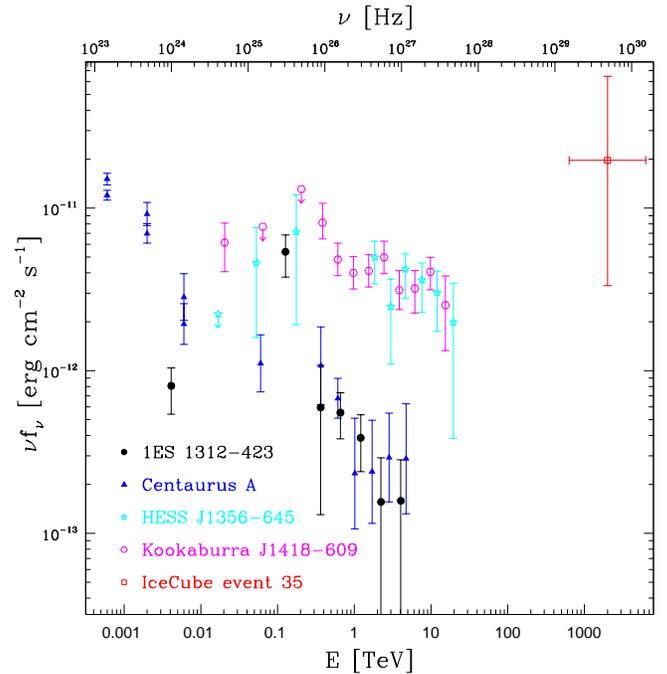}
\caption{$\gamma$-ray SEDs of four sources in the error circle of ID 35, namely: 
the BL Lac 1ES 1312$-$432 (black filled circles) \citep{2013MNRAS.434.1889H}, 
the radio galaxy Centaurus A (blue triangles), and 
the PWNe HESS J1356$-$645 (cyan stars) and 
Kookaburra J1418$-$609 (magenta hexagons) \citep[][and references therein]{2013ApJ...773...77A}. 
The (red) open square represents the neutrino flux for the corresponding IceCube event; error bars as
described in Fig. \ref{SED_MKN421}.}
\label{SED_ID35}
\end{figure}

Fig. \ref{SED_ID33} shows the $\gamma$-ray SEDs of three sources in the error circle of ID 33,
namely 1RXS J193109.5+093714, the strongest BL Lac, MGRO J1908+06, 
the strongest TeV source and a PWN, and HESS J1857+026, an extended Galactic source. 
The latter two both reach $\sim 20 - 40$ TeV. However, only MGRO J1908+06 seems to be 
a plausible astronomical counterpart, since HESS J1857+026 appears to be too soft. 
The BL Lac reaches only $\sim 60$ GeV with an upper limit at $\sim 1$ TeV a factor of 3 below the fluxes of
the two Galactic sources.


Fig. \ref{SED_ID35} shows the $\gamma$-ray SEDs of four sources in the error circle of ID 35, which is
the IceCube event with the largest energy ($\sim 2$ PeV). These include: 1ES 1312$-$432, the only 
blazar in TeVCat, Centaurus A, and two PWNe, HESS J1356$-$645 and 
Kookaburra J1418$-$609. 
The PWNe are more than one order of magnitude brighter than the two extragalactic 
ones at $E > 1$ TeV but still appear to fail the ``energetic'' test. We therefore cannot match the most energetic IceCube event with a plausible
astronomical counterpart. However, as discussed in Sec. \ref{sec:hybrid_bl}, we cannot exclude that
one of the 1FHL sources in the same error box is a TeV emitter and responsible for the neutrino emission. 
Note that in the case of the radio galaxy and the BL Lac the EBL does not 
make much of a difference: 
Centaurus A is at $z = 0.002$ while the correction for 1ES 1312$-$432, which has a redshift in between 
those of PKS 2005$-$489 and H 2356$-$309 ($z=0.105$), will therefore be between a factor $\sim 3$ and 
$\sim 10$ (see Figs. \ref{SED_2005} and \ref{SED_2356}). Interestingly, the predicted neutrino
flux from Centaurus A at the energy of ID 35 is $\approx 1,000$ times smaller 
than the flux connected with that event
\citep{2013APh....48...30S}.


\begin{table*}
\caption{List of most probable counterparts of selected IceCube high-energy neutrinos.}
\begin{tabular}{@{}llll}
IceCube ID  & Counterpart(s) & Class & Catalogue(s)  \\
\hline
#9 &  MKN 421	 &  BL Lac	 (HSP) & TeVCat/WHSP \\
   & 	  1ES 1011+496 & BL Lac	 (HSP) &  TeVCat/WHSP\\
10 & H 2356$-$309	&  BL Lac	 (HSP)	  & TeVCat/WHSP \\
14 &   HESS J1809$-$193 & PWN	 & TeVCat   \\
  17 &  PG 1553+113	& BL Lac	 (HSP)	  & TeVCat/WHSP\\
19 &  1RXS J054357.3$-$553206 &  BL Lac	 (HSP)	  & WHSP\\
20 & 	SUMSS J014347$-$584550 & 	BL Lac	 (HSP)	  & WHSP\\
22 &  1H 1914$-$194 & 	BL Lac	 (HSP)	  & WHSP\\
27 &  PMN J0816$-$1311	 & BL Lac	 (HSP) & WHSP\\
33 &  MGRO J1908+06 &	PWN	 & TeVCat   \\
\hline
\multicolumn{3}{l}{}\\
\end{tabular}
\label{tab:lik_couterparts}
\end{table*}

In Fig. \ref{SED_2356} we plot the $\gamma$-ray SED of NGC 253 as well, a starburst galaxy also 
in the error circle of ID 10. As could be anticipated by its relatively low VHE flux, the SED drops steeply and clearly fails the ``energetic'' diagnostic. Given its
very low redshift ($z=0.0008$) the effect of the EBL is negligible and therefore NGC 253 is an extremely 
unlikely counterpart of the IceCube event. As was the case for Centaurus A, a very recent paper by \cite{2014ApJ...780..137Y} 
predicts for NGC 253 a neutrino flux at the energy of ID 10 $\approx 2,000$ times smaller 
than the flux connected with that event.  
We present our list of most plausible counterparts in Sect. \ref{sec:disc}. 



\section{Discussion}\label{sec:disc}

Our goal was to investigate the origin of the first high-energy astrophysical neutrinos ever detected through a
model independent approach. The most natural sources to begin with are $\gamma$-ray emitters. 
Starting from Tab. \ref{tab:counterparts} and \ref{tab:non_counterparts}, we have studied the photon -- neutrino hybrid 
SEDs of all possible counterparts. This turned out to be a very powerful discriminant for the likelihood
of the association, especially given the relatively  large error circles of the neutrino detections. 

In the comparisons presented in Fig.~\ref{SED_MKN421} to Fig.~\ref{SED_ID35}, 
a gap is present typically from $\approx 10 - 30$ TeV to (neutrino) energies $\ga 100$ TeV (apart 
from Fig. \ref{SED_SUMSS_0143} where the gap is much larger).
Any possible extrapolation to close such a gap is highly non trivial. 
Moreover, strong variability and different emission peak
energies displayed by blazars complicate even further the matter. We have nevertheless
singled out $\gamma$-ray sources, which we believe are plausible counterparts. 

Only 10\% (6/47) of the TeVCat counterparts passed the  ``energetic''  test, the others being 
too $\gamma$-ray faint.
In order to make up for the rather scanty sky coverage of TeVCat, particularly in the high-latitude sky, we 
have also taken into consideration the WHSP blazars still not TeV-detected, applying the ``energetic'' test to 
them as well but considering in this case the strong likelihood that they might be TeV emitters. 
This way we singled out four WHSP-only objects and ranked them as probable counterparts.
The 1FHL-only sources, although in some cases intriguing from the energetic side, turned
out to have too large of a gap between their highest electromagnetic energy and that of the IceCube event
to make a sensible link between photons and neutrinos. Moreover, without even a hint of a possible
emission in the VHE band, we decided not to rank in this work the 1FHL-only sources as probable candidates.

Our findings are summarized in Tab. \ref{tab:lik_couterparts}, which gives the most probable neutrino 
counterparts for 9/18 of the IceCube events. We stress that this does not mean that the remaining nine 
events have no astronomical counterparts but rather that there is less information available for the sources
in their error circles for us to be able to judge on their association likelihood. 

The counterparts turned out to be mostly HSP BL Lacs and two PWNe, 
pushing forward a mixed scenario of Galactic and extra-galactic neutrino sources, so far not yet considered
in the literature. We note that one of the three PeV events (ID 14) appears to have only one plausible 
counterpart, a Galactic one. 
The second PeV IceCube event  (ID 20) has one WHSP BL Lac counterpart but no Galactic ones, 
while the third and most energetic PeV event (ID 35) is at present without likely 
counterparts. NGC 253 (ID 10), which is the only starburst galaxy in our sample,
is too weak when compared to other candidates. Same story for Centaurus A (ID 35), which is the 
only radio galaxy in our sample and an all-time favoured as a potential source by the 
cosmic ray community.

For completeness, we have also checked that our neutrino fluxes are consistent with 
the IceCube upper limits provided by \cite{2013ApJ...779..132A} (see their Tab. 2 and 3)
assuming an $E^{-2}$ ($\nu^{-1}$) spectrum. This is especially important for the northern sources, 
where the neutrino non-detections are particularly stringent and at the level of the fluxes here derived 
(within the rather larger error bars). If the proposed counterparts in Tab. \ref{tab:lik_couterparts} are 
indeed neutrino sources, this suggests that a direct detection by IceCube is within reach. 

To check the significance of our cross-correlations, we have re-done them by 
shifting the positions of the $\gamma$-ray catalogues by varying amounts
always larger than the largest error radius ($\ge 20^{\circ}$). Based on this,
the chance of having seven or more neutrino events associated with TeVCat sources is 
$\sim 64\%$, which reduces to $\sim 50\%$ for five neutrino events and blazar only counterparts. 
This simply reflects the incompleteness of TeVCat\footnote{As shown by taking a random half of the
WHSP sample and obtaining a value of $\sim 55\%$ for four neutrino events, which is much
larger than the probability derived for the whole sample, as detailed here.}. 
As regards WHSP, the chance of having eight neutrino events associated 
with that sample is 4$\%$, while that of having 16 neutrino events associated with 1FHL 
sources is $\sim 4\%$, with the same probability for 1FHL blazars only becoming $\sim 2\%$. 
Even if some of these values are intriguing, 
we are fully aware that, having used the median angular errors for the neutrino events as given by the 
IceCube team provides only lower limits to these numbers.

We stress that, while better statistics are undoubtedly needed on the neutrino side, we have also
shown that more TeV observations reaching also higher energies than currently available 
are badly needed to bridge the gap between photon and neutrino energies.
Moreover, based on the energies and fluxes listed in Tab. \ref{tab:ICE}, the merging of ``classical'' and neutrino 
astronomy will require sensitivities $\approx 4 \times 10^{-11}$ erg cm$^{-2}$ s$^{-1}$ at $E \approx 140$ TeV 
and $\approx 3 \times 10^{-11}$ erg cm$^{-2}$ s$^{-1}$ at $E \approx 1$ PeV.  
We note that, while the former values are within reach of the Cherenkov Telescope Array (CTA) even for 
relatively short 
exposure times \citep[see Fig. 7 of ][]{2013arXiv1307.3409B}, the latter are not, as the maximum energy that 
CTA is expected to reach is only $\sim 100$ TeV. 
Needless to say, all high FoM WHSP sources, but in particular those in Tab. \ref{tab:counterparts}, are obvious
targets for current TeV facilities. 


Further upcoming IceCube observations will be able to confirm or disprove the associations suggested here. 
We can think of two possible future scenarios: 

\begin{itemize}
\item if our conclusions are proven to be wrong, this will mean the following: 1. the existence of a (new?)
population of {\it Fermi}(1FHL)-undetected TeV sources either still to be detected, given the poor sky coverage of 
TeV telescopes, or totally absorbed by the extragalactic background light; in the latter case, the high-energy 
neutrino background will not be resolved into individual sources and will remain {\it diffuse} (in IceCube lingo); 
 2. a complete decoupling between 
$\gamma$-ray photons and neutrino production in the Universe; 3. a mixture of the two scenarios described 
above. 
In all these cases,  the identification of neutrino sources will be extremely challenging;
\item if our conclusions are at least partially confirmed, we will have shown for the first time the 
unambiguous presence of hadronic processes in blazars and pulsar wind nebulae, with very 
important consequences for our understanding of jet and high-energy astrophysics.
\end{itemize}

\section{Conclusions}
\label{subsec:Conclusion}

We have taken a very simple approach to tackle the issue of the astronomical
counterparts of the IceCube neutrinos by looking for high-energy ($\ga 10$
GeV) sources within the error circles of the events. We found 
counterparts for sixteen out of the eighteen events considered in this work. 
By studying their hybrid photon -- neutrino SEDs we have narrowed down our search
and come up with a list of most probable counterparts for nine IceCube neutrinos. 
Interestingly, there is no single class of sources, which we can connect the IceCube 
events with. Instead, the available data suggest a mixed scenario of Galactic and extra-galactic 
neutrino sources. These include BL Lacs of the HSP type (i.e. with peak of the synchrotron 
emission at $\nu > 10^{15}$ Hz) and pulsar wind nebulae. The still TeV-undetected sources, which 
we have singled out as probable neutrino counterparts, are obvious candidates for detection by current
TeV facilities. CTA should reach the sensitivities required for a merging of ``classical'' and neutrino 
astronomy at $\approx 100$ TeV. 
Further upcoming IceCube observations 
will be able to confirm or disprove the associations suggested here. 
 Whatever the outcome of these tests, this will have crucial implications for blazar jets, high-energy 
astrophysics, and cosmic-ray and neutrino astronomy.

\section*{Acknowledgments}
We thank Paolo Giommi, Stefan Coenders, Luigi Costamante, Andreas Gross, and Sirin Odrowski 
for useful discussions and suggestions and the many teams, 
which have produced the data and catalogues used in this paper for making this work possible. 
E. R. is supported by a Heisenberg Professorship of the Deutsche Forschungsgemeinschaft (DFG RE 2262/4-1).
We acknowledge the use of data and software facilities from the ASDC, managed
by the Italian Space Agency (ASI). This research has made use of the VizieR
catalogue access tool, CDS, Strasbourg, France and of the NASA/IPAC
Extragalactic Database (NED), which is operated by the Jet Propulsion
Laboratory, California Institute of Technology, under contract with the
National Aeronautics and Space Administration.



\label{lastpage}
\end{document}